\documentclass[acmsmall]{acmart}
\AtBeginDocument{%
  }

\usepackage{multirow}
\usepackage{fontawesome}
\usepackage{hyperref}
\usepackage{algorithm}
\usepackage{algorithmic}
\usepackage[most]{tcolorbox}
\usepackage{xcolor}
\newtcolorbox{AnswerBox}{
  colback=gray!20,
  colframe=black,
  arc=2mm,
  boxrule=1.0pt,
  left=6pt,
  right=6pt,
  top=6pt,
  bottom=6pt,
}

\setcopyright{acmlicensed}
\copyrightyear{2025}
\acmYear{2025}
\acmDOI{XXXXXXX.XXXXXXX}
\acmISBN{978-1-4503-XXXX-X/2018/06}

\begin{document}

\title{InfCode: Adversarial Iterative Refinement of Tests and Patches for Reliable Software Issue Resolution}


\author{Kefan Li}
\affiliation{%
  \institution{Beihang University}
  \city{Beijing}
  \country{China}
}
\affiliation{%
  \institution{Beijing Tokfinity Technology Co., Ltd.}
  \city{Beijing}
  \country{China}
}

\author{Mengfei Wang}
\author{Hengzhi Zhang}
\author{Zhichao Li}
\affiliation{%
  \institution{Beijing Tokfinity Technology Co., Ltd.}
  \city{Beijing}
  \country{China}
}

\author{Yuan Yuan}
\author{Mu Li}
\author{Xiang Gao}
\author{Hailong Sun}
\author{Chunming Hu}
\author{Weifeng Lv}
\affiliation{%
  \institution{Beihang University}
  \city{Beijing}
  \country{China}
}

\renewcommand{\shortauthors}{Li et al.}

\begin{abstract}
Large language models have advanced software engineering automation, yet resolving real-world software issues remains difficult because it requires repository-level reasoning, accurate diagnostics, and strong verification signals. Existing agent-based and pipeline-based methods often rely on insufficient tests, which can lead to patches that satisfy verification but fail to fix the underlying defect.
We present \textbf{InfCode}, an adversarial multi-agent framework for automated repository-level issue resolution. InfCode iteratively refines both tests and patches through adversarial interaction between a Test Patch Generator and a Code Patch Generator, while a Selector agent identifies the most reliable fix. The framework runs inside a containerized environment that supports realistic repository inspection, modification, and validation.
Experiments on SWE-bench Lite and SWE-bench Verified using models such as DeepSeek-V3 and Claude 4.5 Sonnet show that InfCode consistently outperforms strong baselines. 
It achieves \textbf{79.4\%} performance on SWE-bench Verified, establishing a new state-of-the-art. 
We have released InfCode as an open-source project at \url{https://github.com/Tokfinity/InfCode}.
\end{abstract}

\maketitle

\section{Introduction}
Large language models (LLMs) have recently demonstrated strong capabilities across many software engineering tasks. 
Their ability to interpret natural language specifications, generate executable code, and interact with external tools has motivated increasing interest in automating software issue resolution. 
This task requires diagnosing the behavioral discrepancy described in an issue report and synthesizing a correct code modification that integrates into a complex repository. 
Software issue resolution refers to identifying the root cause of a reported bug and producing an appropriate code modification that restores the intended behavior, ensure the reliable of software systems \cite{freund1997decision,jimenez2023swe,guo2025omnigirl}. 
Compared with isolated code tasks, resolving repository-level issues requires a more comprehensive understanding of multi-file dependencies, project-specific invariants, and execution behaviors. 

LLMs have achieved impressive results at the function level in code generation \cite{chen2021evaluating,austin2021program,li2022competition}, code repair \cite{olausson2023self,shinn2023reflexion}, and test generation \cite{kang2023large, nie2023learning}, yet repository-level issue resolution remains substantially more challenging. 
To mitigate these difficulties, two main lines of work have emerged. 
Agent-based approaches such as SWE-agent \cite{yang2024swe}, OpenHands \cite{wang2024openhands}, Moatless-Tools \cite{orwall2024moatless}, AutoCodeRover \cite{zhang2024autocoderover}, SpecRover \cite{ruan2024specrover}, and Trae-Agent \cite{gao2025trae} provide LLMs with tool interfaces for repository exploration, including file editing, shell execution, and code searching. 
These systems benefit from flexible and adaptive reasoning, as agents can iteratively gather information and validate intermediate hypotheses. 
Pipeline-based methods, such as Agentless \cite{xia2024agentless}, decompose the task into structured subtasks with targeted prompting, which improves stability and reduces error accumulation during reasoning.

Recent systems often attempt to generate tests that reproduce the reported issue, moving beyond reliance on existing repository tests. 
While this represents an important advancement, the generated tests are frequently not sufficiently strong, not accurately aligned with the issue semantics, or unable to capture subtle behavioral constraints. 
As a result, agents may optimize code patches against weak or imperfect tests and thus fail to fully resolve the underlying problem. 
Moreover, most prior work treats test generation and code generation as loosely connected steps. 
Without a mechanism that explicitly coordinates the two, the verification signal remains limited and may permit patches that only satisfy partial or over-simplified test conditions. 

To address these limitations, we propose \textbf{InfCode}, an adversarial multi-agent framework for automated repository-level issue resolution. 
The framework introduces two specialized agents that engage in iterative adversarial refinement. 
A Test Patch Generator constructs and strengthens test cases based on the issue description, aiming to expose the incorrect behavior more effectively. 
A Code Patch Generator responds to these strengthened tests by producing improved code modifications. 
This adversarial interaction drives both agents toward more rigorous testing and more robust patches. 
A Selector agent subsequently evaluates all candidate patches and identifies the most reliable one. 
All agents operate inside a containerized environment that ensures reproducible execution and provides a suite of tools for repository inspection, modification, and validation.

We evaluate InfCode on SWE-bench Lite \cite{jimenez2023swe} and DeepSeek-V3 \cite{liu2024deepseek} and further test it with stronger models such as Claude 4.5 Sonnet \cite{anthropic_claude_sonnet4.5_2025} on SWE-bench Verified \cite{openai_swe_bench_verified_2024}. 
The results show consistent improvements over strong agent-based and pipeline-based baselines and establish a new state-of-the-art (SOTA) (79.4\%) on SWE-bench Verified. 
Ablation studies confirm the effectiveness of adversarial iteration and the importance of the final selection stage.
This demonstrates that adversarially iterating between test generation and code patching provides a principled and effective strategy for enhancing the reliability and generality of automated software issue resolution.
In summary, we propose the following contributions:
\begin{itemize}
\item We propose an adversarial multi-agent framework that iteratively refines both code and test patches through generation and selection, leading to higher patch quality and robustness.
\item We develop a repository-aware tool suite that enhances agents’ understanding and manipulation of real-world codebases within a containerized environment.
\item We conduct comprehensive experiments showing that our approach outperforms existing automated patch generation methods in both correctness and generalization.
\end{itemize}

\section{Method}
\subsection{Framework Overview}
\begin{figure}[h]
\centering
\includegraphics[width=\linewidth]{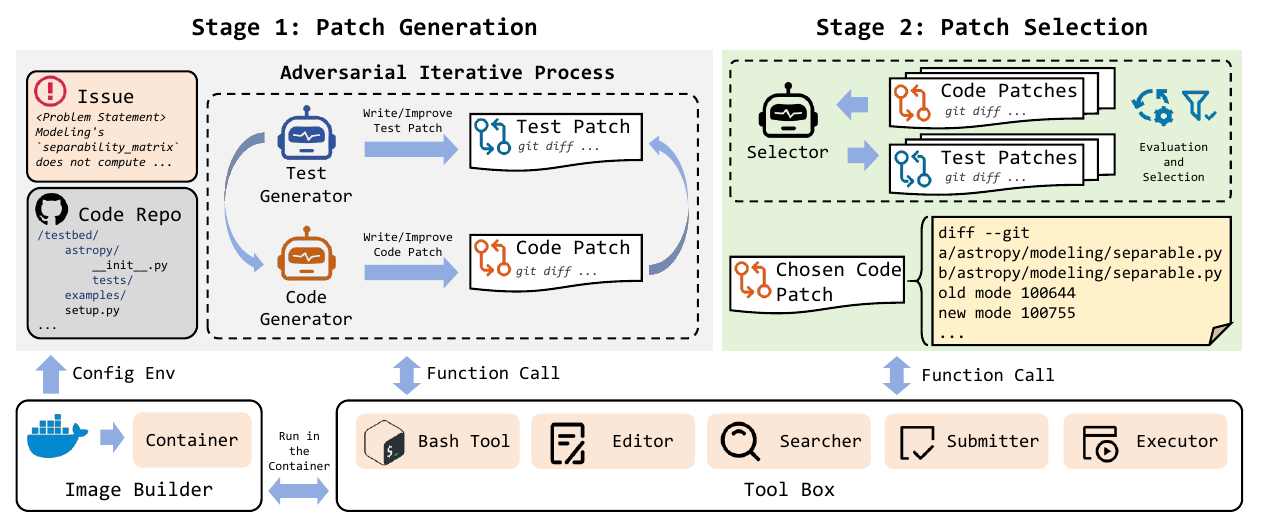}
\caption{
Overview of InfCode for automated code patch generation and selection.
}
\label{fig:framework}
\Description{
Overview of InfCode for automated code patch generation and selection.
}
\end{figure}

Our framework adopts a multi-agent architecture that automates software patch generation and validation through adversarial collaboration between agents. As illustrated in Figure~\ref{fig:framework}, the overall process is divided into two stages: Stage 1: Patch Generation and Stage 2: Patch Selection. 
The system begins with the automatic construction of a controlled environment using a containerized image builder, followed by the orchestration of specialized tools, such as bash utilities, code editors, searchers, submitters, and executors, that support the end-to-end patch generation and evaluation workflow.

At the core of the framework lies a dual-agent adversarial generation mechanism, where the Code Patch Generator Agent and the Test Patch Generator Agent interact iteratively to produce and refine code patches. The primary motivation for this design is that existing test cases often fail to expose the problematic behavior described in a reported issue. Therefore, we introduce an adversarial loop in which the Test Generator continuously strengthens its test suite, while the Code Generator incrementally refines the code to satisfy increasingly stringent requirements. This interplay ensures that the final code patch is not only functional but also robust against a comprehensive set of test conditions.

By combining containerized reproducibility, multi-agent cooperation, and adversarial iteration, the framework enables the automatic synthesis of high-quality code patches that effectively resolve reported issues while maintaining software reliability.

\subsection{Stage 1: Patch Generation}

Stage 1 focuses on producing candidate patches through an \textbf{adversarial iterative process} between the Test Generator and the Code Generator. Given an issue and its associated code repository, the Test Generator first creates test patches that attempt to reproduce the faulty behavior. The Code Generator then produces code patches aimed at passing these newly generated tests. Both agents operate inside the controlled container environment, invoking necessary operations through the tool suite.

\paragraph{Adversarial Iterative Refinement.}
The core innovation of this stage lies in the adversarial iterative refinement mechanism between code and test generation. 
When the Code Generator produces a patch that successfully passes the current test suite, the Test Generator re-analyzes the issue to identify weaknesses, missing edge cases, or insufficient coverage. 
It then introduces additional or stronger test cases that challenge the existing implementation. 
In response, the Code Generator must refine its code to satisfy these enhanced tests. 
This iterative competition and cooperation between the two agents leads to a dynamic equilibrium where both the tests and code progressively improve: tests become increasingly comprehensive and discriminative, while code becomes more robust and generalizable. 
Ultimately, this adversarial generation strategy encourages the emergence of high-quality patches that address the issue thoroughly and resist regression under unseen conditions. 
To prevent unbounded iteration between the Test Generator and the Code Generator, we impose a hard cap on the number of iterations, once this limit is reached, the generation process terminates immediately. 
In addition, if the Test Generator strengthens the test suite and the code still passes all tests, the refinement loop terminates at once.

\subsection{Stage 2: Patch Selection}
After a series of candidate patches are produced in Stage 1, Stage 2 focuses on identifying the optimal code patch through systematic evaluation and selection. The Selector agent collects all generated code patches along with their associated test patches and evaluates them based on predefined performance metrics such as functional correctness, test coverage, execution success, and compatibility with repository constraints.

The Selector may execute the candidate patches within the containerized environment to verify their behavior empirically. It then compares their performance, filtering out overfitted or unstable solutions. Ultimately, the Selector selects the most reliable and generalizable code patch—the one that not only resolves the reported issue but also passes the strengthened test suite produced during adversarial iteration. This chosen patch represents the final output of the framework and can be directly integrated into the target repository.

\subsection{Tool Suite and Execution Environment}
To ensure reproducibility and consistent execution, all operations in our framework are performed within a Docker container instantiated for each issue. The container provides an isolated environment where a suite of tools collaboratively supports patch generation, editing, and evaluation. The following tools are integrated within the containerized workflow:

\begin{itemize}
\item \textbf{Bash Tool:} Executes shell commands within the container, enabling task automation such as running tests, managing dependencies, and invoking build scripts.
\item \textbf{Editor:} Supports file creation, content insertion and replacement, and retrieval of specific line ranges, allowing fine-grained source code modification.
\item \textbf{Searcher:} Performs efficient iterative directory searches using \texttt{ripgrep}, with support for regular expressions to locate relevant code or test fragments.
\item \textbf{Submitter:} Runs \texttt{git diff} to extract and submit patch contents, maintaining version control consistency and recording patch provenance.
\item \textbf{Executor:} Acts as the interface between tools and the Docker container, forwarding execution requests and managing input/output interactions.
\end{itemize}

This tool suite collectively enables automated, reproducible, and isolated code patch generation and evaluation within the adversarial multi-agent framework.

\section{Experimental Setup}

\subsection{Research Questions}
We propose the following research questions (RQs):
\begin{itemize}
\item \textbf{RQ1:} How well does the proposed agent framework perform in generating correct and robust patches compared with existing methods?
\item \textbf{RQ2:} How does each component of the framework contribute to its overall effectiveness?
\item \textbf{RQ3:} What is the maximum performance achieved by InfCode when applied to SOTA LLMs?
\end{itemize}

\subsection{Benchmarks}

We conducted the experiments of RQ1 using SWE-bench Lite \cite{jimenez2023swe}, a lightweight subset of the SWE-bench dataset, which is particularly suitable for lightweight evaluations. 
Many baselines have been evaluated on this subset. 
To optimize resource usage, we employed the DeepSeek-V3 model for the experiments. 
DeepSeek-V3 \cite{liu2024deepseek} possesses strong code generation and tool invocation capabilities, making it an ideal choice as the backbone LLM for agent systems. 
Additionally, to assess the absolute performance of our method, we evaluated it on the SWE-bench Verified \cite{openai_swe_bench_verified_2024} subset and the highly capable model Claude 4.5 Sonnet \cite{anthropic_claude_sonnet4.5_2025}. 
The SWE-bench Verified subset is a curated version of the SWE-bench dataset, where both solutions and tests have undergone rigorous manual validation. 
The tests in this subset are sufficiently rigorous, enabling a more stringent evaluation of patches. 
Our method achieved SOTA performance on the SWE-bench Verified leaderboard.

\subsection{Metrics}
We employed the evaluation method provided by the official SWE-bench.
After generating the patches, we used the \verb|git diff| command to extract them. 
It is important to note that, in order to avoid interference from modifications to test files, we excluded all files starting with \verb|test| from the evaluation. 
Subsequently, we conducted the evaluation using the SWE-bench command-line interface. 
We reported the solved rate, the number of problems solved, and the average cost per problem (in USD).

\subsection{Baselines}
On SWE-bench Lite, we compared various baselines that performed well on this dataset. 
The solved rate and cost data were obtained from their respective papers/reports or the SWE-bench Lite leaderboard. 
To ensure a fair comparison, we primarily selected experimental data from models such as DeepSeek-V3 \cite{liu2024deepseek} and GPT-4o \cite{openai2024gpt4o}. 
On SWE-bench Verified, we compared our method with the top five performing methods, with the data sourced from the SWE-bench Verified leaderboard.

\section{Results}

\subsection{RQ1: Overall Effectiveness Comparison}

\paragraph{Performance on SWE-bench Lite.}
Table~\ref{tab:rq1} presents the performance of InfCode compared to other baseline methods on the SWE-bench Lite dataset. 
We observe that, among the 300 problems in the SWE-Bench Lite dataset, InfCode with Deepseek-V3 as the backbone model successfully solves 118 problems. 
This performance is the best among baselines utilizing similar models. 
Furthermore, although our method incurs higher costs than certain baselines, such as the KGCompass, it outperforms them. 
Additionally, thanks to the relatively low API cost of the Deepseek-V3 model, our approach significantly reduces costs while improving performance, compared to methods using GPT-4.

\begin{table}[h!]
\caption{
Comparison of InfCode and other baseline methods on the SWE-bench Lite dataset.
}
\label{tab:rq1}
\begin{tabular}{llcc}
\toprule
\multirow{2}{*}{\textbf{Method}} & \multirow{2}{*}{\textbf{LLM}} & \multicolumn{2}{c}{\textbf{SWE-bench Lite}} \\
\cmidrule{3-4}
&& \textbf{Resolved (\%)} & \textbf{Avg. Cost (\$)} \\
\midrule
SWE-agent \cite{yang2024swe} & GPT-4o & 18.33\% (55) & 2.53 \\
OpenHands \cite{wang2024openhands} & GPT-4o & 22.00\% (66) & 1.72 \\
Agentless \cite{xia2024agentless} & GPT-4o & 32.00\% (96) & 0.70 \\
SpecRover \cite{ruan2024specrover} & Sonnet-3.5+GPT-4o & 31.00\% (93) & 0.65  \\
AutoCodeRover \cite{zhang2024autocoderover} & GPT-4 & 19.00\% (57) & 0.43  \\
& GPT-4o & 30.67\% (92) & -     \\
SWE-Search \cite{antoniades2024swe} & GPT-4o & 31.00\% (93) & -     \\
Moatless Tools \cite{orwall2024moatless} & DeepSeek-V3 & 30.67\% (92) & - \\
& GPT-4o & 24.67\% (74) & - \\
KGCompass \cite{ma2025thinking} & DeepSeek-V3 & 36.67\% (110) & 0.20 \\
\textbf{InfCode} & \textbf{DeepSeek-V3} & \textbf{40.33\% (121)} & 0.26 \\
\bottomrule
\end{tabular}
\end{table}

\begin{figure}[h!]
\centering
\includegraphics[width=0.8\linewidth]{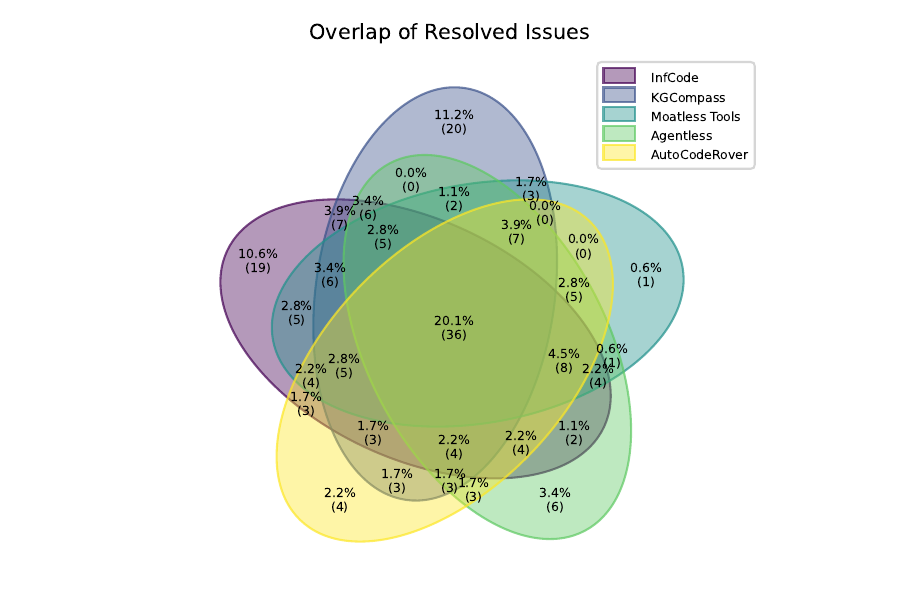}
\caption{
Illustrate the overlap between issues resolved by InfCode and the top baseline methods.
}
\label{fig:venn}
\Description{
Illustrate the overlap between issues resolved by InfCode and the top baseline methods.
}
\end{figure}

\paragraph{Unique issues fixed.}
Figure~\ref{fig:venn} illustrates the problem coverage and differences between InfCode and the baseline methods. 
Among the problems solved, 36 are common to all methods. 
In addition, among all the methods, InfCode and KGCompass solved the most unique problems. 
Furthermore, InfCode solved 11 more problems than KGCompass.
This highlights the superior performance of the InfCode method.

\begin{AnswerBox}
Answer RQ1: On the SWE-bench Lite dataset, using the DeepSeek-V3 model and models of similar capabilities, InfCode solved the most problems and also addressed the highest number of unique problems, highlighting the superior performance of the InfCode method.
\end{AnswerBox}

\subsection{RQ2: Ablation Study of Framework Components}

To investigate the contribution of each module in InfCode, we conducted ablation experiments. 
Specifically, we removed the Adversarial Iteration and Selection modules, resulting in the configurations ``w/o Adversarial'' and ``w/o Selection'', and performed experiments on the SWE-bench Lite dataset using the DeepSeek-V3 model. 
The results are presented in Table~\ref{tab:rq2}.

\begin{table}[h!]
\caption{
Ablation study of InfCode on the SWE-bench Lite dataset using DeepSeek-V3.
}
\label{tab:rq2}
\begin{tabular}{lc}
\toprule
\textbf{Method} & \textbf{Resolved (\%)} \\
\midrule
\textbf{InfCode} & \textbf{40.33\% (121)} \\
w/o Adversarial & 36.33\% (109) \\
w/o Selection   & 32.33\% (97)  \\
\bottomrule
\end{tabular}
\end{table}

The results demonstrate that both modules contribute to the performance of InfCode, with the removal of either module leading to a decrease in performance. 
Notably, the Selection module has a greater impact on the overall performance than the Adversarial Iteration module. 
This highlights the effectiveness of both the adversarial iteration and patch selection strategies.

\begin{AnswerBox}
Answer RQ2: Both the Adversarial Iteration and Selection modules contribute to the performance of InfCode, with the Selection module making a larger contribution.
\end{AnswerBox}

\subsection{RQ3: Performance on SOTA LLMs}

Although the performance on SWE-bench Lite is promising, it may suffer from a potential issue of insufficient test intensity, which could lead to an inability to filter out potentially incorrect patch repairs. Additionally, the current DeepSeek-V3 \cite{liu2024deepseek} model is not the strongest model, and performance on the most advanced models would be more persuasive. 
Therefore, we also conducted experiments on the SWE-bench Verified \cite{openai_swe_bench_verified_2024} dataset using the strongest model, Claude 4.5 Sonnet \cite{anthropic_claude_sonnet4.5_2025}. 
The results are presented in Table~\ref{tab:rq3}.

\begin{table}[h!]
\caption{Comparison of InfCode (based on the Claude 4.5 Sonnet) with the top $5$ methods on the SWE-bench Verified leaderboard as of November 14, 2025.}
\label{tab:rq3}
\begin{tabular}{lcc}
\toprule
\textbf{Method} & \textbf{Resolved (\%)} & \textbf{Rank} \\
\midrule
\faTrophy \space \textbf{InfCode} & \textbf{79.4\% (397)} & \textbf{1} \\
TRAE + Doubao-Seed-Code & 78.80\% (394) & 2 \\
Atlassian Rovo Dev (2025-09-02) & 76.80\% (384) & 3 \\
EPAM AI/Run Developer Agent & 76.80\% (384) & 3 \\
v20250719 + Claude 4 Sonnet \\
ACoder & 76.40\% (382) & 5 \\
Warp & 75.60\% (378) & 6 \\
\bottomrule
\end{tabular}
\end{table}

In Table~\ref{tab:rq3}, we report the performance of InfCode on Claude 4.5 Sonnet, along with the performance of the top five methods, excluding InfCode, on the current SWE-bench Verified Leaderboard. 
The results show that InfCode achieved an impressive pass rate, securing the top position. 
This further confirms the effectiveness of InfCode. 

\begin{AnswerBox}
Answer RQ3: On the SWE-bench Verified dataset, InfCode with Claude 4.5 Sonnet as the backbone model achieves the top performance, demonstrating the effectiveness of InfCode.
\end{AnswerBox}

\section{Discussion}

\begin{figure}[h!]
\centering
\includegraphics[width=1.0\linewidth]{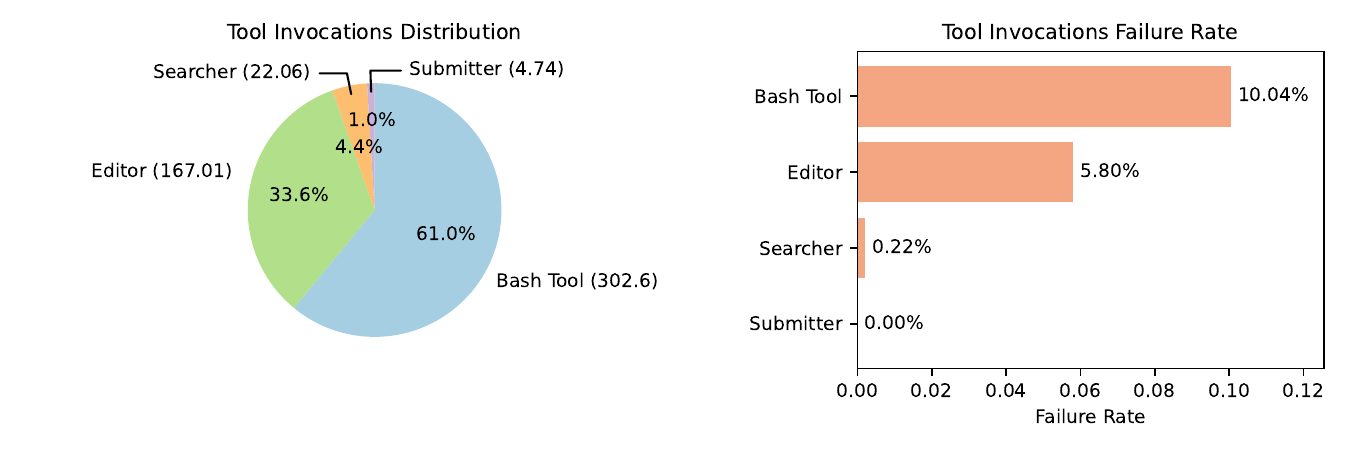}
\caption{
Distribution of the average number of tool invocations per problem and the corresponding failure rates.
}
\label{fig:tools}
\Description{
Distribution of the average number of tool invocations per problem and the corresponding failure rates.
}
\end{figure}

In this section, we analyze the failure rate of tool invocations by LLMs. 
The system provides four tools that can be called directly by the model: Bash Tool, Editor, Searcher, and Submitter. 
We collected statistics on the average number of calls and the failure rates of these tools when InfCode solves a problem. 
The distribution of call counts for each tool is shown in the left panel of Figure~\ref{fig:tools}, and the failure rates for each tool are reported in the right panel of Figure~\ref{fig:tools}.

From the distribution of tool invocations, InfCode calls the Bash Tool most frequently, followed by the Editor. 
This indicates that most attempts focus on modifying files and executing code. 
InfCode performs extensive file inspection, code editing, and debugging, and these operations contribute substantially to successful problem solving. 
The next most common is the Searcher, which primarily supports initial problem localization, once the relevant files are identified, the Editor is used to inspect their contents. 
The Submitter is invoked least often and is used only when the model judges that the problem has been solved.
From the failure rate plot, the Bash Tool exhibits the highest failure rate. 
Inspection of execution logs shows that the main causes include attempts to run complex scripts with \verb|python -c "xxx"| and attempts to execute nonexistent commands, largely due to unfamiliarity with the environment or available commands. 
The Editor has the next highest failure rate, mostly because its string replacement function requires an exact \verb|old_str|, which can be challenging in some contexts. 
Overall, the failure rates across tools remain low, which demonstrates the robustness of InfCode.

\section{Limitations}
Although InfCode attains strong performance, several limitations and avenues for improvement remain. 
First, prior works \cite{ahmed2024tdd,ahmed2025otter,mundler2024swt} have shown that LLM-based agents may produce incorrect test patches when reproducing issues, which can misdirect subsequent code repair. 
Our log analysis reveals a similar pattern: during adversarial iterations, the Test Generator sometimes creates specialized tests that deviate from the issue description in order to induce failures, which in turn misleads the Code Generator. 
This suggests that strategies are still needed to improve the accuracy and faithfulness of the Test Generator. 
Second, the Bash Tool and the Editor continue to exhibit some invocation errors. 
The implementations of these tools require further refinement, and the usage guidelines should be clarified to reduce failure rates as much as possible.

\section{Threads to Validity}
\paragraph{Internal Validity.}
One potential internal threat arises from the possibility that the generated patches might inadvertently modify the corresponding test files, which could bias the evaluation results. Such modifications may artificially increase the apparent success rate of generated patches by weakening or bypassing test cases. To mitigate this risk, we explicitly exclude all files with names beginning with “test” when computing the patch differences using the \texttt{git diff} command. This ensures that only genuine code modifications are considered in the evaluation, preserving the integrity and fairness of the results.

\paragraph{External Validity.}
A potential external threat concerns the generalizability of our approach to programming languages other than those evaluated in our experiments. While our current study focuses on a specific language environment, the proposed framework itself is designed to be language-agnostic. It operates primarily through regular-expression-based search, command-line execution, and repository-level operations driven by bash commands. Consequently, with minor adaptations to language-specific syntax and build tools, the framework can be readily extended to other programming languages and ecosystems.

\paragraph{Construct Validity.}
Construct validity relates to whether our evaluation metrics and experimental setup accurately reflect the true effectiveness of automated patch generation. A possible concern is that passing existing or generated test cases may not fully represent real-world correctness or semantic equivalence to developer fixes. To address this, we evaluate patches not only by test success but also by robustness under adversarially strengthened test suites and by cross-validation against multiple independent test generations. This design ensures that our measurements align with the intended construct of patch quality and that the evaluation outcomes genuinely reflect functional correctness and generalizability.

\section{Related Work}
\subsection{Repository-Level Issue Resolution}
Repository-level software issue resolution has emerged as a central challenge in leveraging LLMs for real-world development workflows. 
The introduction of SWE-bench \cite{jimenez2023swe} provided an executable benchmark grounded in authentic GitHub issues with verifiable test-based evaluation, highlighting a substantial gap between the impressive surface-level generation capabilities of LLMs and the robustness required for practical software maintenance. 
Subsequent studies have shown that even reported ``solved'' issues are frequently incorrect or incomplete \cite{wang2025solved}, underscoring the need for both reliable evaluation and improved reasoning mechanisms.

To bridge this gap, multiple efforts explore agent-based approaches for repository navigation, reasoning, and patch synthesis. 
Frameworks such as SWE-agent \cite{yang2024swe}, OpenHands \cite{wang2024openhands}, and MarsCode \cite{liu2024marscode} offer general-purpose agent execution platforms, while systems like AutoCodeRover \cite{zhang2024autocoderover} and CodeR \cite{chen2024coder} introduce multi-agent or role-specialized collaboration guided by task graphs or iterative retrieval. 
LingmaAgent \cite{ma2025alibaba} extends repository exploration to include historical commits, pull requests, and dependency relationships, and Lingma SWE-GPT \cite{ma2024lingma} integrates development-process-centric knowledge to improve coherence with real engineering workflows. 
Meanwhile, AGENTLESS \cite{xia2024agentless} questions the necessity of complex agent orchestration, demonstrating that reducing step-wise error propagation can improve repair reliability.

Another line of work focuses on enhancing repository understanding and structured reasoning. 
RepoGraph \cite{ouyang2024repograph} and repository-aware knowledge graph approaches \cite{yang2025enhancing} inject structural code relationships, while SpecRover \cite{ruan2024specrover} targets intent extraction to improve semantic grounding. 
Search-enhanced methods such as SWE-Search \cite{antoniades2024swe} employ Monte Carlo Tree Search for guided repair, and competitive strategies like SWE-Debate \cite{li2025swe} use multi-agent argumentation to refine solutions. 
Experience-driven systems, including SWE-Exp \cite{chen2025swe} and EXPEREPAIR \cite{mu2025experepair}, incorporate memory mechanisms to transfer prior solution strategies across repositories.

Recent findings further demonstrate that compute scaling during inference, rather than simply model size, is crucial for successful repository-level repair. 
Thinking Longer, Not Larger \cite{ma2025thinking} and Trae Agent \cite{gao2025trae} show that increased test-time reasoning depth and reflection significantly improve performance. 
Complementary work such as BugPilot \cite{sonwane2025bugpilot} focuses on generating complex bug scenarios to better train and stress-test these systems.

\subsection{Iterative Code–Test Feedback}
Iterative refinement driven by execution feedback has emerged as a crucial paradigm for improving the correctness and reliability of LLM-generated code. 
Early work on model self-debugging \cite{chen2023teaching} demonstrated that LLMs can leverage compilation or runtime error messages to analyze failures and propose targeted revisions, establishing the principle that feedback grounded in actual program behavior is essential for meaningful correction. 
This idea was further generalized in Self-Refine \cite{madaan2023self}, which introduced a task-agnostic iterative feedback loop in which model outputs are repeatedly critiqued and refined. 
Building upon these concepts, Reflexion \cite{shinn2023reflexion} incorporates verbal self-reflection to enable agents to accumulate reasoning strategies across iterations, while RLEF \cite{gehring2024rlef} employs reinforcement learning from execution performance signals to directly shape the model's repair policy.

A complementary direction enhances the feedback loop by involving both code and tests. LLMLOOP \cite{ravi2025llmloop} and CoCoEvo \cite{li2025cocoevo} jointly generate tests and candidate patches, fostering co-evolution of validation and repair, whereas LEVER \cite{ni2023lever} emphasizes verification-in-the-loop, using constraint checks and invariant validation to filter incorrect solutions. 
More recent work revisits self-debugging through self-generated tests \cite{chen2025revisit}, demonstrating that adaptive test synthesis substantially improves fault localization. 
Meanwhile, RepairAgent \cite{bouzenia2024repairagent} operationalizes iterative feedback within an autonomous agent framework, integrating code navigation, environment interaction, and repeated refinement to solve real repair tasks.

Beyond system design, theoretical analyses show that iterative code repair inherently involves an exploration–exploitation trade-off \cite{tang2024code}, where excessive refinement may lead to local minima, while insufficient refinement may fail to converge. 
Together, these works converge on the insight that effective code repair relies not only on initial generation quality, but also on structured, feedback-driven iterative improvement tightly aligned with executable program semantics.

\section{Conclusion}
This paper introduced InfCode, an adversarial multi-agent framework designed to improve the reliability and robustness of automated repository-level issue resolution. The framework integrates iterative collaboration between a Test Patch Generator and a Code Patch Generator, enabling tests to be progressively strengthened while patches are refined to satisfy increasingly rigorous requirements. This adversarial co-evolution produces verification signals that more accurately capture the true semantics of reported issues. In addition, the Selector agent evaluates candidate patches within a controlled containerized environment, ensuring consistent execution, reducing instability, and identifying the most reliable solution.

Extensive experiments on SWE-bench Lite and SWE-bench Verified demonstrate that InfCode achieves substantial improvements over existing agent-based and pipeline-based approaches. With Claude 4.5 Sonnet as the backbone model, InfCode attains a solve rate of 79.4\% on SWE-bench Verified, which represents the current state of the art. Ablation studies further highlight the contributions of adversarial iteration and patch selection, confirming that both components are essential for maximizing system performance.
Overall, InfCode shows that coupling adversarial test refinement with iterative patch generation provides an effective strategy for advancing automated software repair. Future research may explore applying this framework to additional programming languages and development ecosystems, enhancing the intelligence of test generation, and integrating learned heuristics to guide interaction between agents more efficiently.


\bibliographystyle{ACM-Reference-Format}
\bibliography{paper}

\appendix

\end{document}